\documentclass{article}

\PassOptionsToPackage{numbers, compress}{natbib}

\usepackage[preprint]{neurips_2025}

\usepackage[utf8]{inputenc} 
\usepackage[T1]{fontenc}    
\usepackage{hyperref}       
\usepackage{url}            
\usepackage{booktabs}       
\usepackage{amsfonts}       
\usepackage{nicefrac}       
\usepackage{microtype}      
\usepackage{xcolor}         

\usepackage{amsmath}
\usepackage{graphicx}
\usepackage[capitalize,noabbrev]{cleveref}
\usepackage[most]{tcolorbox}
\usepackage{multirow}
\usepackage{colortbl}
\usepackage{threeparttable} 
\usepackage{tabularx} 
\usepackage{array} 
\usepackage{enumitem}
\setlist[itemize]{leftmargin=0.2in}
\usepackage{wrapfig}
\usepackage{comment}

\newcolumntype{C}{>{\centering\arraybackslash}X}
\newcolumntype{L}{>{\raggedright\arraybackslash}X} 

\definecolor{msg_generator_color}{RGB}{184,146,48}
\colorlet{msg_generator_color_fill}{msg_generator_color!5!white}
\definecolor{reflector_color}{RGB}{176,36,24}
\colorlet{reflector_color_fill}{reflector_color!5!white}

\title{Learn as Individuals, Evolve as a Team: Multi-agent LLMs Adaptation in Embodied Environments}

\author{%
  Xinran Li\textsuperscript{1,2}
\quad Chenjia Bai\textsuperscript{2,}\thanks{Corresponding authors.}
\quad Zijian Li\textsuperscript{1}
\quad Jiakun Zheng\textsuperscript{2,3}
\quad Ting Xiao\textsuperscript{3}
\quad Jun Zhang\textsuperscript{1,*} \\
\textsuperscript{1}The Hong Kong University of Science and Technology \\
\textsuperscript{2}Institute of Artificial Intelligence (TeleAI), China Telecom \\
\textsuperscript{3}East China University of Science and Technology \\
\texttt{xinran.li@connect.ust.hk, baicj@chinatelecom.cn, } \\
\texttt{zijian.li@connect.ust.hk, \{zhengjk@mail,xiaoting@\}ecust.edu.cn,} \\
\texttt{eejzhang@ust.hk}
}
\begin{document}

\maketitle

\begin{abstract}
    Large language models (LLMs) possess extensive knowledge bases and strong reasoning capabilities, making them promising tools for complex, multi-agent planning in embodied environments. However, despite LLMs' advanced abilities and the sophisticated modular design of agentic methods, existing LLM-based planning algorithms remain limited by weak adaptation capabilities to multi-agent embodied scenarios. We address this limitation by introducing a framework that enables LLM agents to learn and evolve both before and during test time, equipping them with environment-relevant knowledge for better planning and enhanced communication for improved cooperation. Inspired by centralized training with decentralized execution in multi-agent reinforcement learning, we propose a \textit{Learn as Individuals, Evolve as a Team (LIET)} paradigm for multi-agent LLMs adaptation. At the individual level, LLM agents learn a local utility function from exploratory datasets to better comprehend the embodied environment, which is then queried during test time to support informed decision-making. At the team level, LLM agents collaboratively and iteratively maintain and update a shared cooperation knowledge list based on new experiences, using it to guide more effective communication. By combining individual learning with team evolution, LIET enables comprehensive and flexible adaptation for LLM agents. Our experiments on Communicative Watch-And-Help and ThreeD-World Multi-Agent Transport benchmarks demonstrate that LIET, instantiated with both LLaMA and GPT-4o, outperforms existing baselines and exhibits strong cooperative planning abilities. \looseness=-1
\end{abstract}

\section{Introduction} \label{sec: intro}





Multi-agent decision-making is a ubiquitous challenge in real-world applications ranging from daily tasks like completing household chores to industrial-scale problems in domains such as supply chain management~\citep{MAS_SCM_lee2008multi,MAS_SCM_giannakis2011multi}, smart cities~\citep{MAS_building_cai2016general,MAS_trafic_chu2019multi,MARL_autonomous_driving} and robot swarms~\citep{marl_robot_swamy2020scaled}. These problems are inherently challenging due to their long-horizon sequential nature and the complex interaction patterns among agents. \looseness=-1

The emergence of large language models (LLMs)~\citep{llama_touvron2023llama,Llama3_dubey2024llama,gpt4_achiam2023gpt,gpt4o_hurst2024gpt,deepseekr1_guo2025deepseek} has opened new opportunities for multi-agent decision-making. With their strong reasoning capabilities and extensive knowledge priors, LLMs have been employed as agent planners, proposing actions based on the current state of the environment. Recent works~\citep{CoELA_zhangbuilding, ProAgent_zhang2024proagent, roco_mandi2024roco} have demonstrated the potential of prompting LLMs to achieve effective collaboration in long-horizon planning tasks in embodied environments. LLM-based approaches often achieve strong zero-shot performance with their inherent comprehension, reasoning, and planning abilities. Despite these advantages, directly employing off-the-shelf LLMs as planners can be suboptimal without proper adaptation. LLMs, while trained on vast text-based data, lack the abilities to fully comprehend embodied contexts or facilitate dynamic multi-agent cooperation — both essential for effective multi-agent planning in embodied environments. This raises a key question: \textit{can multi-agent LLMs adapt to such tasks by learning and evolving through interactions with their environment and peers, rather than relying exclusively on zero-shot abilities?}

Much like humans who adapt to new environments and teammates over time, LLM-based agents can improve by learning from experience. By interacting with the environment and other agents, they can refine their understanding of the physical world and enhance collaborative behaviors. To this end, we propose a novel LLM-based multi-agent decision-making framework: \textit{Learn as Individuals, Evolve as a Team (LIET)}. LIET enables LLM-based agents to adapt to specific multi-agent embodied planning tasks through model fine-tuning and prompt evolution, enhancing both individual decision-making and team collaboration. LIET augments LLM-powered planners with environment-specific knowledge and collaboration skills derived from both prior experiences and test-time interactions. Inspired by the centralized training and decentralized execution (CTDE) paradigm from multi-agent reinforcement learning (MARL), our framework employs a semi-centralized approach where agents first explore individually to learn environment-specific information, and then collaborate as a team through communication and reflection to achieve better outcomes. To instantiate the \textit{Learn as Individuals, Evolve as a Team} framework, we make the following technical contributions: \looseness=-1
\begin{itemize}
    \item \textbf{Utility-guided individual adaptation}: To enable LLM-powered planners to adapt to specific environmental contexts and make more informed decisions within limited time budgets, we augment their inputs with a utility function that estimates the cost of candidate plans. This utility function is implemented by finetuning a small LLM with a LoRA adapter and a value head on exploratory datasets. \looseness=-1
    \item \textbf{Evolving prompting for communication}: To enhance team adaptation and collaboration, we design a dynamic prompting mechanism for constructive communication during test-time. This mechanism maintains a knowledge list of effective communication strategies that evolves over time, being iteratively refined through a reflection module at the receiver side. The module evaluates whether received messages improve planning outcomes in new situations, thereby fostering increasingly effective multi-agent communication. \looseness=-1
    \item \textbf{Superior performance}: Through extensive experiments on multi-agent planning benchmarks, including Communicative Watch-And-Help (C-WAH)\citep{CWAH_puigwatch, CoELA_zhangbuilding} and ThreeDWorld Multi-Agent Transport (TDW-MAT)\citep{TDW_gan2022threedworld, CoELA_zhangbuilding}, we demonstrate that LIET significantly outperforms existing baselines. Our results highlight its effectiveness in enabling long-horizon multi-agent collaborative planning. \looseness=-1
\end{itemize}

\vskip -0.2in
\section{Background}

\subsection{Problem Setting}
We consider a fully cooperative multi-agent decision-making problem, which can be characterized as a decentralized partially observable Markov decision process (Dec-POMDP)~\citep{pomdp_oliehoek2016concise}. The Dec-POMDP is formalized by a tuple $\mathcal{M} = \langle N, T, \mathcal{S}, \mathcal{O}, \mathcal{A}, P, \mathcal{G}, R \rangle$, where $N$ is the number of agents and $T$ is the task episode horizon. $\mathcal{S}$ represents the set of global states, and $\mathcal{O} = \mathcal{O}^{\text{env}} \times \mathcal{O}^{\text{comm}} $ is the set of local observation, consisting of partial environmental state information and messages from other agents. The action set is composed of high-level actions $\mathcal{A} = \mathcal{A}^{\text{env}} \cup \mathcal{A}^{\text{comm}}$, where $\mathcal{A}^{\text{env}}$ represents actions executed in the embodied environment and $\mathcal{A}^{\text{comm}}$ involves broadcasting natural language messages to other agents. The transition probability $P(s' \mid s, \boldsymbol{a})$ defines the likelihood of reaching a new state $s' \in \mathcal{S}$ after taking joint actions $\boldsymbol{a} = [ a_1, a_2, \ldots, a_N \mid a_i \in \mathcal{A}, i \in \{ 1, 2, \ldots, N \}  ]$ in the current state $s \in  \mathcal{S}$. $g \in \mathcal{G}$ specifies the common goal for all agents to achieve, and the reward $r = R(s)$ represents the degree of completeness of the team goal $g$.  \looseness=-1

In our experiments, we instantiate this problem in long-horizon embodied environments, assigning agents household tasks $g \in \mathcal{G}$ such as \texttt{Transport 2 bananas and 1 apple to the bed} or \texttt{Prepare a meal}. The agents can perform high-level actions $a \in \mathcal{A}^{\text{env}}$ such as \texttt{Grab(object)} or \texttt{GoTo(location)}, but execution failures may occur if actions are not well-situated in the current state of the environment (e.g., attempting to grab an object that is not present).

\subsection{Related Work}
\paragraph{LLM-based Single-agent Planning}
Recent advancements in large language models (LLMs) have led to the development of highly capable models, including LLaMA 3~\citep{llama_touvron2023llama,Llama3_dubey2024llama}, GPT-4~\citep{gpt4_achiam2023gpt,gpt4o_hurst2024gpt}, and DeepSeek-R1~\citep{deepseekr1_guo2025deepseek}. By leveraging their extensive general knowledge and exceptional reasoning capabilities, LLM-based agents have achieved remarkable success in embodied decision-making tasks that integrate perception, action, and interaction with physical or simulated environments~\citep{reasoning_survey_sun2023survey}.
Seminal works such as SayCan~\citep{saycan_brohan2023can}, grounded decoding~\citep{grounded_decoding_huang2024grounded} and Text2Motion~\citep{text2motion_lin2023text2motion} ground agent actions by using value functions to bridge language with real-world tasks. Follow-up studies have enhanced LLM planning capabilities by introducing thinking modules~\citep{inner_mono_huang2023inner,swiftsage_lin2024swiftsage} or world abstractions~\citep{world_nottingham2023embodied}. Others have leveraged in-context learning~\citep{few_shot_LLMPlanner_song2023llm} when expert demonstrations are available. Another research direction focuses on aligning visual perception with language objectives, resulting in algorithms such as VLMaps~\citep{VLMaps_huang2023visual}, TaPA~\citep{Tapa_wu2023embodied}, mobility VLA~\citep{mobility_VLA_xumobility}, and NavGPT~\citep{navgpt_zhou2024navgpt,NavGPT2_zhou2024navgpt}. These methods refine LLMs’ ability to plan and act in environments requiring seamless integration of language and vision. \looseness=-1

\paragraph{Multi-agent Collaboration in Embodied Environments} 
Following the success of LLM-based single-agent planning, researchers have sought to extend these capabilities to multi-agent systems, aiming to improve task efficiency~\citep{cooperative_survey_torreno2017cooperative,coopertive_survey_guo2024large}. Early approaches adopt a decentralized perspective, using modular designs~\citep{mindagent_gong2024mindagent} to facilitate cooperation by introducing communication modules~\citep{CoELA_zhangbuilding} or teammate belief modules~\citep{ProAgent_zhang2024proagent,ToM_li2023theory}. While these decentralized approaches ensure flexibility, performance may suffer due to the lack of global consensus, particularly in tasks requiring a high degree of collaboration~\citep{cent_dec_chen2024scalable}. To address this, centralized methods such as RoCo~\citep{roco_mandi2024roco} and CaPo~\citep{capo} incorporate multi-round discussions among agents to enhance collaboration. However, as task complexity increases, partial observability becomes significant; or as the number of agents grows, these methods struggle to maintain consensus efficiently. Communication costs also scale rapidly, making centralized approaches less practical in real-world scenarios.
Our proposed LIET strikes a balance between decentralized and centralized methods by adopting a semi-centralized framework. It incorporates an individual planning module enhanced with a utility function and facilitates cooperation through an evolving communication scheme that leverages past experiences. This approach enables agents to dynamically tailor their messages to better accommodate complex environments. \looseness=-1

Furthermore, some recent research seeks to address the gap between simulation environments and real-world settings by considering more realistic scenarios. Notably, LLaMAR~\citep{llamar_nayaklong} restricts agent observations to imperfect knowledge from the simulator and employs a verifier to ensure the feasibility of plans, while SMART-LLM~\citep{smarlllm_kannan2024smart} converts high-level instructions into executable actions and deploys these methods on real robots. While these efforts address challenges that are orthogonal to the focus of our work, they could potentially be combined with LIET to enhance its applicability in real-world tasks. \looseness=-1

\section{Method} \label{sec: method}

\begin{figure*}[ht]
\begin{center}
\centerline{\includegraphics[scale=0.58]{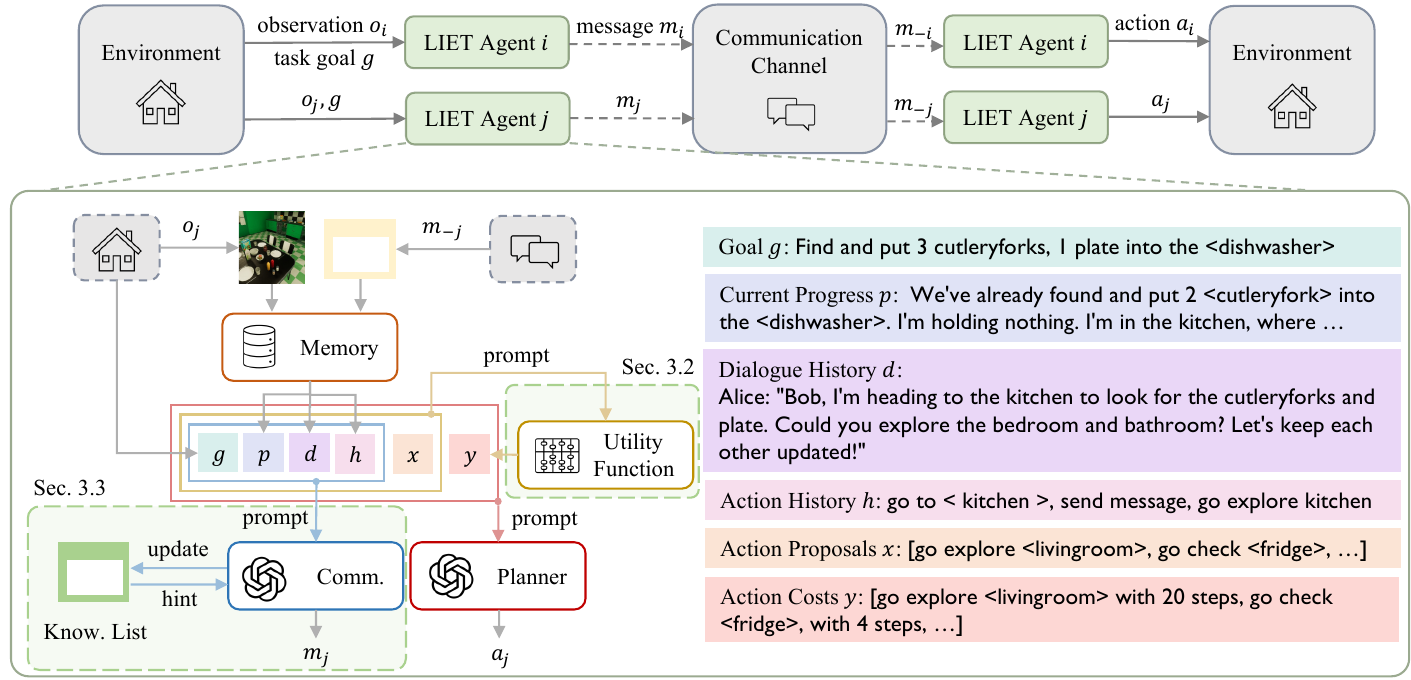}}
\caption{The overall framework of LIET. It adopts a semi-centralized decision making scheme with communication, where each agent plans independently and communicates with others at critical timesteps. During local planning, LIET agents utilize information from the environment, memory, and a utility function (detailed in \cref{sec: learn_ind}) to make informed decisions. For communication, LIET agents generate messages based on the environment, memory, and an iteratively updated knowledge list (abbreviated as Know. List), which contains cooperation tips derived from prior experiences (detailed in \cref{sec: envolve_team}). In LIET, the utility function, communication module (abbreviated as Comm.), and planner are all powered by LLMs.}
\label{fig:framework}
\end{center}
\vskip -0.3in
\end{figure*}

In this section, we propose a novel framework for multi-agent embodied planning, called \textit{Learn as Individuals, Envolve as a Team (LIET)}. LIET adapts multi-agent LLM-based planners to embodied environments by addressing decision-making challenges from both local and global perspectives.
In \cref{sec: framework}, we first examine the challenges of multi-agent embodied planning tasks and analyze how to comprehensively address these challenges by bridging local decision-making with global goals. 
We then introduce the overall framework of LIET, which is built around two core components: (1) a decentralized utility module that enables agents to make more informed and efficient decisions independently (\textit{learn as individuals}), and (2) an evolving communication scheme that enhances cooperation and mutual understanding among agents through iterative prompt improvements (\textit{evolve as a team}).
In \cref{sec: learn_ind}, we elaborate on the utility module, which is instantiated as a fine-tuned LLM that provides cost information to augment local planning. In \cref{sec: envolve_team}, we expand on the evolving communication scheme, which leverages test-time improvements based on agents' prior interactions to enhance team collaboration. \looseness=-1

\subsection{Overall Framework of LIET } \label{sec: framework}
LIET is designed to enhance the capabilities of LLM-enabled planning agents by addressing two key challenges in multi-agent embodied tasks:
(1) \textbf{Limited Environment-relevant Knowledge}: LLMs lack inherent knowledge of the physical world, as they are primarily trained on text corpora rather than embodied tasks~\citep{reasoning_survey_sun2023survey}.
(2) \textbf{Collaboration Gaps}: LLMs are not inherently equipped with the experience or mechanisms to collaborate effectively with other agents. As a result, LLM-based planners often generate plausible yet inefficient or inapplicable plans for embodied environments.

To address these challenges, LIET enhances the prompts of LLM-based planners by incorporating task-relevant knowledge and mechanisms specifically designed for multi-agent embodied environments. The overall framework of LIET, as illustrated in \cref{fig:framework}, introduces two key innovations that distinguish it from existing methods. At the individual level, LIET equips each agent with a pretrained utility function that infers the cost of actions, enabling more efficient use of the effective time management and improving the agent’s ability to generate practical and effective plans as a decentralized decision-maker. At the team level, LIET fosters seamless collaboration by maintaining a shared knowledge list containing cooperation tips, which serves as a dynamic resource for guiding constructive communication among agents. This knowledge list evolves iteratively through reflections on past coomunication experiences, allowing agents to continuously refine their coordination strategies over time. For the ease of illustration, we describe the framework using a two-agent setting. However, LIET is not limited to this configuration and can be readily applied to scenarios involving more agents, as demonstrated empirically in \cref{sec:exp_results}. \looseness=-1

\subsection{Learn as Individuals} \label{sec: learn_ind}
In this subsection, we describe how LIET agents learn as individuals to understand the embodied environment and construct a utility function that assists in decentralized decision-making. The utility function plays a critical role in guiding agents by providing task-relevant information. Importantly, the utility function should meet two key criteria: it must (1) provide environment-related insights that aid decision-making, and (2) remain goal-agnostic to accommodate a wide variety of tasks.

\begin{figure*}[t]
\begin{center}
\centerline{\includegraphics[scale=0.62]{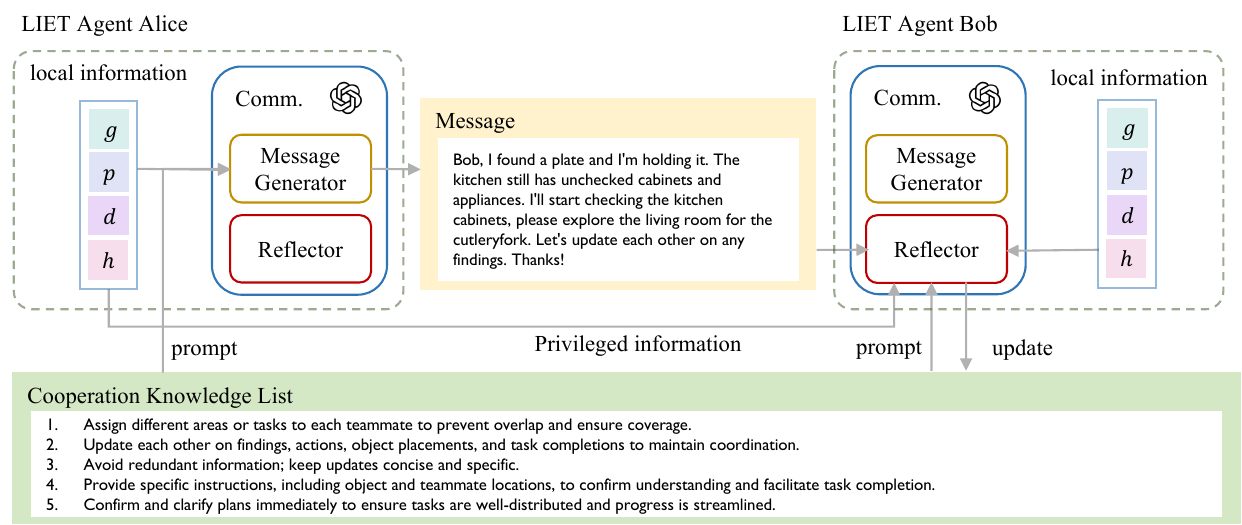}}
\caption{The scheme for LIET communication. In LIET, communication happens at critical timesteps in a broadcast fashion. The key idea is that the LIET agents maintain a common cooperation knowledge list collaboratively based on reflection on previous conversation and use the knowledge to guide future message generation. At a decision timestep, agent Alice generates a message based on local information and the current cooperation knowledge list and sends it to agent Bob. Upon receiving the message, Bob uses it for its local decision making. In the meanwhile, Bob optionally leverages a reflector module and analyzes how the message can be further improved from the receiver's perspective. The insights from such reflection will be merged to the cooperation knowledge list. The example message and the cooperation knowledge list are derived from the CWAH task experiments. The local information contains goal, current progress, dialogue history and action history, as illustrated in \cref{fig:framework}. \looseness=-1}
\label{fig:comm_scheme}
\end{center}
\vskip -0.3in
\end{figure*}

\paragraph{Utility Function to Estimate Action Costs}

Inspired by the affordance function proposed by \citet{saycan_brohan2023can}, we adopt a value-based utility function to estimate the action costs in terms of environmental steps. This cost estimation is crucial for long-horizon embodied planning tasks, as high-level actions often have widely varying execution costs. For instance, traveling to another room may require significantly more steps than placing an object on a tray. By incorporating this information, LIET empowers planners to make more efficient and context-aware decisions. Additionally, the utility function enhances coordination in multi-agent settings. Since agents operate simultaneously in the environment, understanding the cost of their actions helps them allocate efforts more effectively, improving overall task execution. A key advantage of this approach is that collecting cost information does not require task-specific expertise or human expert labeling. Instead, LIET leverages agents' ability in exploring the environment to autonomously gather the necessary data, making it a scalable and efficient solution. \looseness=-1

In LIET, the utility function, denoted as $f(\ell_{o_i}, \ell_a)$, takes the text description of the current local observation $o_i$ and the text description of the action candidate $a$ as input and outputs a scalar value $c = f(\ell_{o_i}, \ell_a)$ representing the estimated number of environmental steps required to execute the action. During test time, LIET agents query the utility function to evaluate the costs of their current action plans, enabling them to make better-informed decisions, as illustrated in \cref{fig:framework}.

\paragraph{Finetuning an LLM as the Utility Function}

Given the LLMs' strong abilities to comprehend the text context, we instantiate the utility function as a finetuned LLM augmented with an additional value head to produce scalar outputs. This design transforms the task into a regression problem, where the objective is to minimize the prediction error on action cost estimation. Specifically, we employ the Mean Squared Error (MSE) loss as the training objective:
\begin{equation}
    \mathcal{L}(\theta) = \mathbb{E}_{\ell_{o_i}, \ell_a, c^{\text{GT}} \sim \mathcal{D}}[(c^{\text{GT}} - f(\ell_{o_i}, \ell_a; \theta) )^2],
\end{equation}
where $\mathcal{D}$ is a pre-collect exploratory dataset consist of description-cost pairs, $c^\text{GT}$ represents the ground-truth cost, and $\theta$ includes the parameters of the LoRA adapter~\citep{lora_hulora} and the value head, which is composed of MLP layers.


\subsection{Evolve as a Team} \label{sec: envolve_team}
While individual agents are equipped with environment-relevant information for more grounded and effective planning, the next challenge lies in adapting LLMs to function as part of a team.
Compared to the individual learning scheme, we adopt a more versatile approach for team cooperation, enabling LIET agents to continuously improve~\citep{era_experience_silver2025welcome} their communication by leveraging incoming experiences during task execution.
The core mechanism of this team-level adaptation is an evolving prompting framework that enhances multi-agent communication. Specifically, agents exchange feedback on the messages they receive and collaboratively summarize these insights into a shared cooperation knowledge list. This knowledge list serves as a guide for generating more constructive messages in future interactions, allowing agents to refine their communication strategies over time. We term this process ``evolve as a team'' because agents collectively contribute to building and improving the shared knowledge during test time. The overall workflow of this approach is illustrated in \cref{fig:comm_scheme}. \looseness=-1

\paragraph{Message Generation Based on Experience}
To enable test-time improvement in multi-agent planning tasks, LIET incorporates prior experience into the message generation process through the shared knowledge list. This list is iteratively updated and used as part of the prompt for message generation, structured as follows:

\begin{tcolorbox}[colframe=msg_generator_color, colback=msg_generator_color_fill, title=Message Generator Prompting]
Prompt: \texttt{<Task Description>} + \texttt{<Local Observation>} + \texttt{<Knowledge List>}. Please help me generate a message ...\\
LLM: \texttt{<Message>}.
\end{tcolorbox}

Here, the placehoder \texttt{<Task Description>} refers to the goal for the task $g$, \texttt{<Local Observation>} contains sender agent's local information such as current progress, action history and dialogue history, \texttt{<Knowledge List>} is the hints maintained by all the agents iteratively on how to generate more constructive messages to teammates, which we will elaborate on in the following paragraph. \looseness=-1

\paragraph{Reflection on the Receiver Side} 


One of the key challenges in developing a test-time evolving scheme for communication lies in identifying effective message exchanges. Communication in multi-agent tasks contributes indirectly to task performance by providing decision-makers (i.e., message receivers) with additional information. However, it is often difficult to assess the relevance or quality of a message during execution.
To address this, LIET introduces a reflection~\citep{reflection_shinn2023reflexion} mechanism that prompts the message receiver to evaluate the usefulness of received messages for decision-making. Since receivers inherently possess more context about what information is helpful for planning, they are better positioned to assess communication effectiveness. After reflecting on the message, the receiver generates insights on how future messages could be improved. These insights are then incorporated into the shared knowledge list, allowing the entire team to benefit from the reflection process. The prompt used for the reflector is as follows:
\begin{tcolorbox}[colframe=reflector_color, colback=reflector_color_fill, title=Reflector Prompting]
Prompt: \texttt{<Task Description>} + \texttt{<Local Observation>} + \texttt{<Privileged Information>} + \texttt{<Received Message>} + \texttt{<Knowledge List>}. Please reflect ...\\
LLM: \texttt{<Updated Knowledge List>}
\end{tcolorbox}

Here, the placeholder \texttt{<Privileged Information>} maintains additional details from the sender, such as their current plans, and the \texttt{<Updated Knowledge List>} replaces the existing \texttt{<Knowledge List>} in the next round of message generation and reflection, ensuring iterative improvement in message quality. \looseness=-1

\section{Experimental Results} \label{sec: experiment}
In this section, we instantiate LIET with different LLMs to test its effectiveness on embodied household cooperative planning tasks. Specifically, we first compare LIET against several baselines to demonstrate its superior performance. Then, we demonstrate the intelligent and cooperative behaviors and analyze the mechanisms behind LIET through visualization and ablation studies. Furthermore, we extend LIET to scenarios with varying numbers of agents, demonstrating the flexibility of its semi-centralized framework. 

\subsection{Environmental Setups}
\paragraph{Benchmark Descriptions} Following previous works~\citep{CoELA_zhangbuilding,capo}, we adopt two long-horizon multi-agent planning benchmarks to evaluate LIET: Communicative Watch-And-Help (C-WAH)~\citep{CWAH_puigwatch,CoELA_zhangbuilding} and ThreeDWorld Multi-Agent Transport (TDW-MAT)~\citep{TDW_gan2022threedworld,CoELA_zhangbuilding} (detailed in \cref{supp: env_details}). Both benchmarks feature cooperative tasks within realistic household simulators, incorporating both language and visual interfaces. Agents can navigate the environment and interact with objects to achieve task goals.

Built on the VirtualHome-Social simulator, C-WAH extends the Watch-And-Help Challenge~\citep{CWAH_puigwatch} to feature multi-agent communication. Agents are required to collaboratively complete household chores by coordinating efforts through communication. The evaluation set consists of 10 episodes spanning five different tasks, each with a horizon of 250 timesteps. Performance is measured as the average timesteps required to complete all tasks.

Based on the ThreeDWorld Transport Challenge~\citep{TDW_gan2022threedworld}, TDW-MAT extends the task to multi-agent settings. Agents must transport objects scattered across different rooms to designated goal locations, using containers to assist in transportation. The evaluation set includes 24 episodes divided into ``food'' and ``stuff'' tasks, with each episode spanning up to 3000 time frames. Performance is evaluated using the transport rate, defined as the percentage of successfully transported objects relative to the total tasked objects within the time budget.

\begin{table}[t]
\caption{Average steps ($\downarrow$) to complete tasks on the C-WAH benchmark under visual (Vis. Obs.) and symbolic (Sym. Obs.) perception. The best-performing method in each setting is highlighted.}
\label{tab:CWAH}
\centering
\makebox[\textwidth]{%
  \begin{threeparttable}
    \begin{tabularx}{1.0\textwidth}{l!{\vrule width \arrayrulewidth}C|CCCC|CCCC}
      \toprule
      \multirow{2}{*}{} &
        \multirow{2}{*}{MHP$^*$} &
        \multicolumn{4}{c|}{LLaMA-3.1 Agents} &
        \multicolumn{4}{c}{GPT-4o Agents} \\ \cmidrule(lr){3-6} \cmidrule(lr){7-10}
        &
        &
        CoELA &
        ProAgent &
        RoCo &
        LIET &
        CoELA &
        ProAgent &
        RoCo &
        LIET \\ \midrule
      Vis. Obs.  &
        103 &
        98.5 &
        103.7 &
        113.9 &
        \cellcolor[gray]{0.9}88.8 &
        93.5 &
        86.2 &
        107.4 &
        \cellcolor[gray]{0.9} 85.0 \\
      Sym. Obs.  &
        75 &
        58.8 &
        64.4 &
        73.9 &
        \cellcolor[gray]{0.9} 50.4 &
        48.4 &
        61.0 &
        64.7 &
        \cellcolor[gray]{0.9}40.3 \\
      \bottomrule
    \end{tabularx}
    \begin{tablenotes}
      \footnotesize
      \item[$*$] MHP results are adopted from the original CoELA paper~\citep{CoELA_zhangbuilding}.
    \end{tablenotes}
  \end{threeparttable}%
}
\vskip -0.1in
\end{table}

\begin{table}[t]
\caption{Transport rate (\%) comparison on the TDW-MAT benchmark tasks with and without oracle perception. Tasks are categorized into ``food'' and ``stuff'' based on the object types. The best-performing method in each setting is highlighted.}
\label{tab:TDW_results}
\centering
\makebox[\textwidth]{%
  \begin{threeparttable}
    \begin{tabularx}{\textwidth}{l|C|CCCC|CCCC}
      \toprule
      \multirow{2}{*}{} &
        \multirow{2}{*}{RHP$^*$} &
        \multicolumn{4}{c|}{LLaMA-3.1 Agents} &
        \multicolumn{4}{c}{GPT-4o Agents} \\ \cmidrule(lr){3-6} \cmidrule(lr){7-10}
        &
        &
        CoELA &
        ProAgent &
        RoCo &
        LIET &
        CoELA &
        ProAgent &
        RoCo &
        LIET \\ \midrule
      \multicolumn{10}{c}{w/o Oracle Perception} \\ \midrule
      Food ($\uparrow$) &
        67 &
        77.5 &
        76.7 &
        75.0 &
        79.2 &
        80.0 &
        82.5 &
        79.2 &
        83.3 \\
      Stuff ($\uparrow$) &
        54 &
        65.0 &
        69.2 &
        69.2 &
        69.2 &
        68.3 &
        70.8 &
        76.7 &
        75.8 \\
      Avg. ($\uparrow$) &
        61 &
        71.3 &
        72.9 &
        72.1 &
        \cellcolor[gray]{0.9} 74.2 &
        74.2 &
        76.7 &
        77.9 &
        \cellcolor[gray]{0.9} 79.6 \\ \midrule
      \multicolumn{10}{c}{w/ Oracle Perception} \\ \midrule
      Food ($\uparrow$) &
        76 &
        84.2 &
        82.5 &
        80.8 &
        82.5 &
        85.0 &
        79.1 &
        85.8 &
        90.0 \\
      Stuff ($\uparrow$) &
        74 &
        75.8 &
        76.7 &
        80.0 &
        78.3 &
        84.2 &
        85.0 &
        85.8 &
        84.2 \\
      Avg. ($\uparrow$) &
        75 &
        80.0 &
        79.6 &
        \cellcolor[gray]{0.9}80.4 &
        \cellcolor[gray]{0.9}80.4 &
        84.6 &
        82.0 &
        85.8 &
        \cellcolor[gray]{0.9}87.1 \\ \bottomrule
    \end{tabularx}
    \begin{tablenotes}
      \footnotesize
      \item[$*$] RHP results are adopted from the original CoELA paper~\citep{CoELA_zhangbuilding}.
    \end{tablenotes}
  \end{threeparttable}%
}
\vskip -0.2in
\end{table}


\paragraph{Baselines}
To evaluate LIET, we benchmark it against two categories of methods: traditional agents and LLM-based agents.

The traditional agents include: (i) \textit{MCTS-based Hierarchical Planner (MHP)}~\citep{CoELA_zhangbuilding}: A hierarchical planning approach designed for the original Watch-And-Help Challenge. It features a Monte Carlo Tree Search (MCTS)-based high-level planner and a regression-based low-level planner. (ii) \textit{Rule-based Hierarchical Planner (RHP)}~\citep{CoELA_zhangbuilding}: A heuristic-based hierarchical planning approach designed for the original ThreeDWorld Transport Challenge. It uses a rule-based high-level planner combined with an A-start-based low-level planner for navigation.

The LLM-based baselines include: (iii) \textit{CoELA}~\citep{CoELA_zhangbuilding}: A decentralized LLM-based planning framework with modular components for perception, communication, planning, and execution. (iv) \textit{ProAgent}~\citep{ProAgent_zhang2024proagent}: A decentralized LLM-based planning method that explicitly models teammate intentions. (v) \textit{RoCo}~\citep{roco_mandi2024roco}: A centralized LLM-based planning scheme that incorporates multi-turn communication for collaborative decision-making.

\paragraph{Implementation Details}

To evaluate LIET across different underlying LLMs, we instantiate the LLM-based planner in LIET and other LLM-based baselines using two state-of-the-art models: LLaMA 3.1-70B\citep{Llama3_dubey2024llama}, an open-source model deployed on the TogetherAI platform, and ChatGPT-4o\citep{gpt4o_hurst2024gpt}, a closed-source model accessed via the OpenAI API. We use temperature = 0.7, top-p = 1, and a maximum token limit of 256 for both models. Unless otherwise stated, all experiments involve two agents on both benchmarks. Further details regarding LIET implementation are provided in \cref{supp: implementation_details}. \looseness=-1

\subsection{Results} \label{sec:exp_results}
\paragraph{Performance}
We summarize the performance of LIET and baseline methods equipped with LLaMA 3.1-70B and ChatGPT-4o on the C-WAH and TDW-MAT benchmarks in \cref{tab:CWAH} and \cref{tab:TDW_results}, respectively. Overall, LIET outperforms the baselines, demonstrating the effectiveness of its semi-centralized framework. By enabling agents to adapt through individual exploration and collaborate via team interactions, LIET significantly enhances performance across both benchmarks.

\begin{figure}[t]
\begin{center}
\centerline{\includegraphics[scale=0.65]{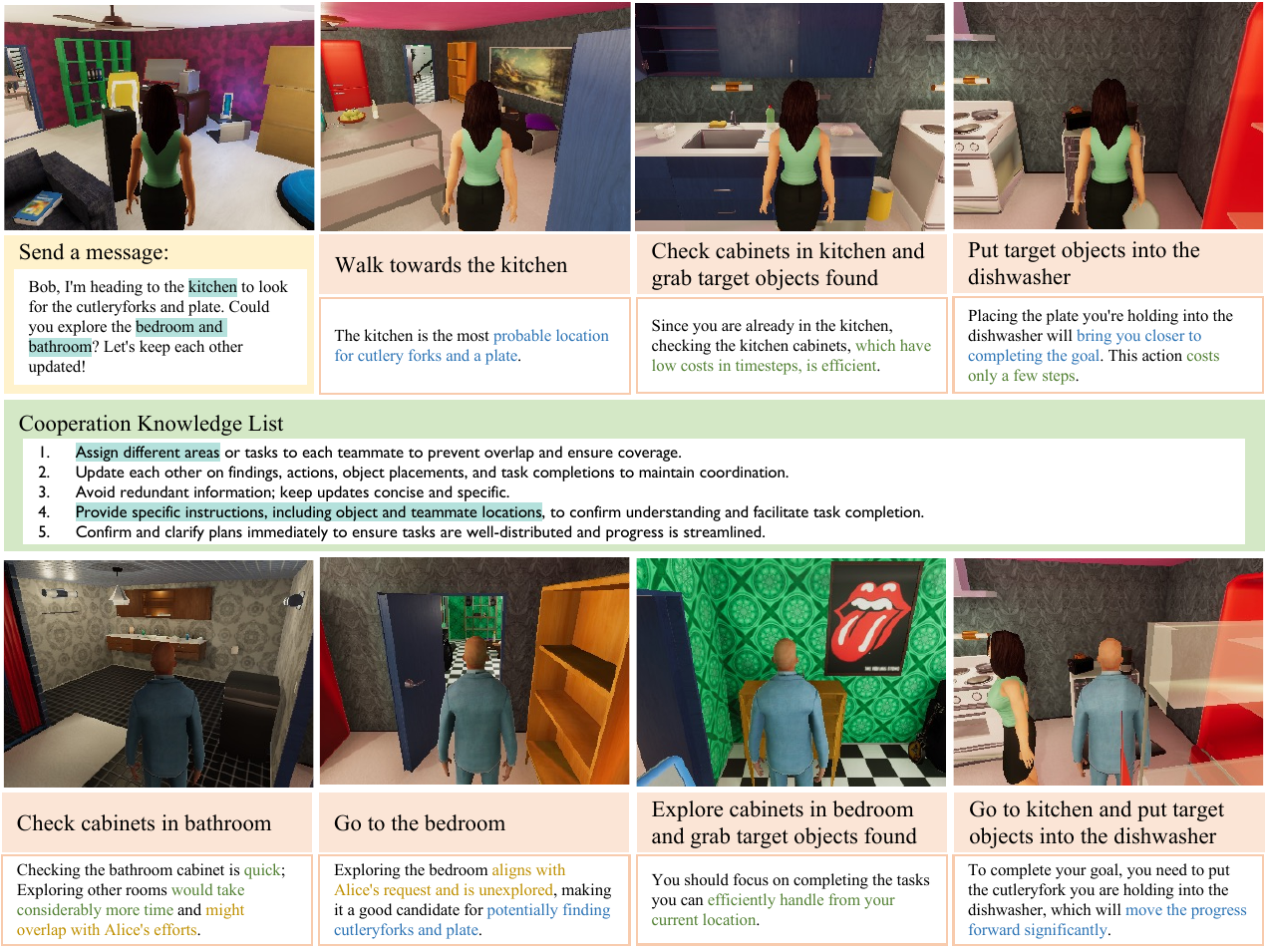}}
\caption{Snapshots of an example episode from the C-WAH benchmark with the task: ``Find and put 3 cutleryforks, 1 plate into the <dishwasher>''. Each agent's visualization includes annotations of their plans and thoughts. The current cooperation knowledge list is also displayed. The message in the yellow dialogue box is directly generated by the LIET agent, while the thoughts are compact excerpts from the agent's Chain of Thought (CoT) process.}
\label{fig:vis_coop}
\end{center}
\vskip -0.3in
\end{figure}

\paragraph{Visualization}
To better understand how LIET operates in practice, we visualize an episode of the C-WAH task in \cref{fig:vis_coop}. Overall, LIET agents exhibit intelligent and cooperative behaviors. Analysis of the agents' CoT reasoning reveals that their decision-making process comprehensively considers three key factors: (1) \textcolor[rgb]{0.2588,0.4510,0.6941}{task progress}, analyzed by the LLMs, (2) \textcolor[rgb]{0.3686,0.5059,0.2471}{efficiency}, measured through costs estimated by the utility function and (3) \textcolor[rgb]{0.7216,0.5725,0.1882}{dialogue history}, generated by the evolving communicative prompting. These considerations enable LIET agents to make informed and considerate decisions. Moreover, the visualization highlights the direct correspondence between the cooperative knowledge list and generated messages, demonstrating how team adaptation guides effective communication. In the visualized episode, agent Alice (in green) initiates the task by proposing a division of labor. Agent Bob (in blue) evaluates Alice's suggestions and selects the more efficient subtask. The two agents then proceed with their respective subtasks, dynamically balancing action costs with the progress made toward task completion. This cooperative strategy exemplifies LIET’s ability to foster efficient collaboration.
Additional visualizations, including scenarios where the agents' exploration areas overlap and more intense discussions are required, can be found in \cref{supp:appendix_more_results}, further demonstrating the communication patterns induced by LIET.



\begin{wraptable}{r}{0.5\textwidth} 
\vspace{-.2in}
\caption{Average steps ($\downarrow$) to complete tasks on the C-WAH benchmark in ablation studies. All agents are powered by GPT-4o.}
\vspace{.1in}
\label{tab:ablation}
\begin{threeparttable}
\begin{tabular}{p{5cm}p{1cm}}
\toprule
\multicolumn{2}{c}{\textbf{LIET Ablations}} \\ \midrule
LIET & 40.3 \\
LIET w/o individual learning & 52.7 \\
LIET w/ prompted costs estimation & 57.6 \\
LIET w/o team evolving & 49.3 \\ \bottomrule
\end{tabular}
\end{threeparttable}
\vspace{-.1in}
\end{wraptable}

\begin{table}[t]
\caption{Average steps ($\downarrow$) to complete tasks on the C-WAH benchmark with symbolic perception, using different numbers of agents. All agents are powered by GPT-4o. The best performing method in each setting is highlighted. }
\label{tab:CWAH_more}
\centering
\makebox[\textwidth]{%
  \begin{threeparttable}
    \begin{tabularx}{0.8\textwidth}{l!{\vrule width \arrayrulewidth}CCCC}
      \toprule
      &
        CoELA &
        ProAgent &
        RoCo &
        LIET \\ \midrule
      2 agents &
        48.4 &
        61.0 &
        64.7 &
        \cellcolor[gray]{0.9}40.3 \\
      3 agents &
        38.9 &
        50.9 &
        44.4 &
        \cellcolor[gray]{0.9}34.9 \\
      4 agents &
        36.5 &
        46.6 &
        53.9 &
        \cellcolor[gray]{0.9}33.1 \\
      \bottomrule
    \end{tabularx}

  \end{threeparttable}%
}
\vskip -0.1in
\end{table}

\paragraph{Ablation Studies} 
We conduct ablation studies on the C-WAH benchmark with symbolic perception to evaluate the impact of key design components in LIET. The results, summarized in \cref{tab:ablation}, compare the full LIET framework with three ablated versions: (i) \textit{LIET w/o individual learning}, in which we remove the utility function instantiated by a finetuned LLM described in \cref{sec: learn_ind}, eliminating individual-level adaptation.  (ii) \textit{LIET w/ prompted costs estimation}, in which we replace the finetuned utility function with an explicit prompt requesting the planner to estimate action costs, simulating a non-adapted cost assessment mechanism.  (iii) \textit{LIET w/o team evolving}, in which we omit the teammate reflection process during test time (detailed in \cref{sec: envolve_team}) and prompt the communication with a fixed template, removing team-level adaptation. The performance degradation across all ablations demonstrates that both individual learning and the team evolving process are critical to LIET’s effectiveness.
These results highlight that successful multi-agent planning with LLMs requires integrated adaptation at both individual and team levels to achieve optimal outcomes. \looseness=-1

\paragraph{LIET with More Agents} 
We further examined the performance of LIET under varying numbers of agents, using GPT-4o as the LLM planner on the C-WAH benchmark. Results in \cref{tab:CWAH_more} demonstrate that LIET consistently outperforms baselines across different agent counts. This is primarily due to LIET’s flexible framework, which integrates local costs estimation (via the pretrained individual utility function) with global cooperation (via team-evolving communication prompts). Notably, the utility functions used across all multi-agent settings were finetuned exclusively on datasets collected from two-agent scenarios. This demonstrates LIET's strong generalization capabilities, showing how adaptation learned from simpler contexts can effectively transfer to more complex multi-agent environments. \looseness=-1

\section{Conclusions} \label{sec: conclusion}
In this work, we explored multi-agent planning tasks in embodied environments and proposed the \textit{Learn as Individuals, Evolve as a Team (LIET)} framework as an adaptation scheme for multi-agent LLMs. LIET significantly enhances planning performance by enabling LLM agents to adapt to specific environments through both individual learning and team evolution.
Our semi-centralized approach implements adaptation at two critical levels: individually, agents learn utility functions during exploration that inform decision-making at test time; as a team, agents employ an evolving prompting scheme that continuously refines communication strategies through mutual feedback. Experiments on multi-agent embodied planning benchmarks demonstrated LIET's superior performance and robustness when extended to additional agents. These results highlighted the importance of adaptation mechanisms over relying solely on LLMs' zero-shot capabilities in complex embodied scenarios. \looseness=-1

While effective, LIET shares limitations with previous approaches regarding multi-modal information processing, relying on pretrained vision modules to convert visual information into text for LLM planning. The comprehensive integration of multi-modal information in embodied environments remains an important direction for future research. \looseness=-1

\bibliographystyle{ACM-Reference-Format} 
\bibliography{bibliography.bib}

\newpage
\appendix
\section{Experimental Details}
\subsection{Implementation Details on LIET} \label{supp: implementation_details}
\paragraph{Utility Functions Finetuning Details} 

For the exploratory dataset, we collected 10 episodes for C-WAH benchmark and 24 episodes for TDW-MAT benchmark. We used 80\% of this data as the training set to finetune a LLaMA 3.2-1B model with LoRA adapters and MLP layers as the value head. The architectural and training hyperparameters are listed in \cref{tab: hyper}.

\begin{table}[th]
    \centering
    \caption{Hyperparameters for utility functions. }
    \label{tab: hyper}
    \begin{tabular}{cc}
    \toprule
    Hyperparameter & Value \\
    \midrule
    Number of MLP layers & $2$ \\ 
    MLP hidden dimension & $32$ \\ 
    \hline \noalign{\vskip 2pt}
    LoRA target modules & [q\_proj, v\_proj] \\
    LoRA rank ($r$) & $8$ \\ 
    LoRA alpha ($\alpha$) & $32$ \\ 
    LoRA dropout rate & $0.1$ \\ 
    Batch size  & $64$ \\ 
    Weight decay & $0.1$ \\ 
    Training epochs & $20$ \\ 
    \bottomrule
    \end{tabular}
\end{table}

\paragraph{Prompt Templates} We provide example prompt templates for the LLM-based modules used in our framework for the C-WAH benchmark in the following.

\begin{center}
\begin{tcolorbox}[breakable,
    colback=gray!10, 
    colframe=black, 
    width=0.95\textwidth, 
    arc=1mm, 
    boxrule=0.5pt,
    title=Prompt for the planner module in C-WAH,
    fonttitle=\bfseries\color{white},
    colbacktitle=gray!70,
    coltitle=white
]
I'm \$AGENT\_NAME\$. I'm in a hurry to finish the housework with my teammate \$OPPO\_NAME\$ together. Given our shared goal, dialogue history, and my progress and previous actions, please help me choose the best available action to achieve the goal as soon as possible. \\
Note that I can hold two objects at a time and there are no costs for holding objects. All objects are denoted as <name> (id), such as <table> (712). 

Goal: \$GOAL\$ \\
Progress: \$PROGRESS\$ \\
Dialogue history: \\
Alice: "Hi, I'll let you know if I find any goal objects and finish any subgoals, and ask for your help when necessary." \\
Bob: "Thanks! I'll let you know if I find any goal objects and finish any subgoals, and ask for your help when necessary." \\
\$DIALOGUE\_HISTORY\$ \\
Previous actions: \$ACTION\_HISTORY\$ \\
Available actions: \$AVAILABLE\_ACTIONS\$ \\
Answer:
\end{tcolorbox}
\end{center}

\begin{center}
\begin{tcolorbox}[breakable,
    colback=gray!10, 
    colframe=black, 
    width=0.95\textwidth, 
    arc=1mm, 
    boxrule=0.5pt,
    title=Prompt for the message generator module in C-WAH,
    fonttitle=\bfseries\color{white},
    colbacktitle=gray!70,
    coltitle=white
]
I'm \$AGENT\_NAME\$. I'm in a hurry to finish the housework with my friend \$OPPO\_NAME\$ together. Given our shared goal, dialogue history, and my progress and previous actions, please help me generate a short message to send to \$OPPO\_NAME\$ to help us achieve the goal as soon as possible. \\
Note that I can hold two objects at a time and there are no costs for holding objects. All objects are denoted as <name> (id), such as <table> (712). 

Goal: \$GOAL\$ \\
Progress: \$PROGRESS\$ \\
Previous actions: \$ACTION\_HISTORY\$ \\
Dialogue history: \\
Alice: "Hi, I'll let you know if I find any goal objects and finish any subgoals, and ask for your help when necessary." \\
Bob: "Thanks! I'll let you know if I find any goal objects and finish any subgoals, and ask for your help when necessary." \\
\$DIALOGUE\_HISTORY\$ \\
Here are some hints to help you generate more useful messages based on previous experiences: \\
\$KNOWLEDGE\_LIST\$ \\
Note: The generated message should be accurate, helpful and brief. Do not generate repetitive messages and please output the message to send only. \\
Message:
\end{tcolorbox}
\end{center}

\begin{center}
\begin{tcolorbox}[breakable,
    colback=gray!10, 
    colframe=black, 
    width=0.95\textwidth, 
    arc=1mm, 
    boxrule=0.5pt,
    title=Prompt for the reflector module in C-WAH,
    fonttitle=\bfseries\color{white},
    colbacktitle=gray!70,
    coltitle=white
]
I'm \$AGENT\_NAME\$. I'm in a hurry to finish the housework with my teammate \$OPPO\_NAME\$ together. \\
Note that I can hold two objects at a time and there are no costs for holding objects. All objects are denoted as <name> (id), such as <table> (712). 

Given a new piece of decision making experience based on our shared goal, my progress, previous actions, and our current plans, please reflect on the dialogue history and update the knowledge list to better guide future effective communication in cooperative planning. 

Goal: \$GOAL\$ \\
Progress: \$PROGRESS\$ \\
Dialogue history: \\
Alice: "Hi, I'll let you know if I find any goal objects and finish any subgoals, and ask for your help when necessary." \\
Bob: "Thanks! I'll let you know if I find any goal objects and finish any subgoals, and ask for your help when necessary." \\
\$DIALOGUE\_HISTORY\$ \\
Previous actions: \$ACTION\_HISTORY\$ \\
Current plans: \$CURRENT\_PLANS\$ \\
Knowledge list: \\
------------------------- \\
\$KNOWLEDGE\_LIST\$ \\ 
------------------------- \\
Note: Please help update the knowledge list within the '-----'. The updated list should integrate insights from both new experiences and previously accumulated knowledge, aiming to provide hints to enhance future communication message exchange between teammates. This knowledge list is shared among all teammates to enable effective communication to improve decision-making during our decentralized planning. Keep the list concise and informative, not exceeding 100 words.
\end{tcolorbox}
\end{center}

\paragraph{Codebase} We conduct our experiments using the following codebase:
\begin{itemize}
    \item CoELA~\citep{CoELA_zhangbuilding}: \url{https://github.com/UMass-Embodied-AGI/CoELA}
\end{itemize}

We implement the baselines following:
\begin{itemize}
    \item ProAgent~\citep{ProAgent_zhang2024proagent}: \url{https://github.com/PKU-Alignment/ProAgent}
    \item RoCo~\citep{roco_mandi2024roco}: \url{https://github.com/MandiZhao/robot-collab}
\end{itemize}

The code for this work will be released publicly after the review process.

\subsection{Additional Details on Environments} \label{supp: env_details}

We follow the evaluation schemes established in previous work~\citep{CoELA_zhangbuilding,capo}. For completeness, we include details of the experimental environments below.

\subsubsection{TDW-MAT}
\paragraph{Tasks}
TDW-MAT features two distinct tasks: the \textit{food-transporting} and \textit{stuff-transporting} challenges. Each task involves different target objects and container types. The agents' objective is to transport maximum target objects to designated goal positions, utilizing containers as transportation tools. Each container can accommodate up to three objects simultaneously, whereas agents without containers are limited to carrying only two objects at once. Success is measured by the number of objects successfully transported within a 3000-frame time limit.

\paragraph{Layouts} 
The environment comprises 4 floorplans, each with 3 layout variations—two floorplans serve as training environments while the remaining two are reserved for testing.

The food-transporting task requires agents to collect 6 types of edible items (apple, banana, orange, bread, loaf bread, and burger) using 3 available containers (bowl, plate, and tea tray). Meanwhile, the stuff-transporting task involves gathering 6 types of personal items (calculator, mouse, pen, lighter, purse, and iPhone) using different carriers (plastic basket, wood basket, and wicker basket). Each task scenario contains 10 target objects and between 2-5 containers distributed throughout the environment.

The virtual space consists of four room types—living room, office, kitchen, and bedroom—with objects placed according to real-world expectations. For instance, food items are predominantly found in kitchen areas, while stuff items are typically located in office spaces.

\paragraph{Observation Space} Agents in TDW-MAT receive these observations:
\begin{itemize}
    \item \textbf{RGB image}: the egocentric image comes from the camera facing forward, with screen size 512 × 512 and field of view 90; 
    \item \textbf{Depth image}: the depth image has the same camera intrinsic parameters as the RGB image
    \item \textbf{Oracle Perception} (optional): an image where each object id is mapped to a color and the camera intrinsic parameters are the same as the RGB image; 
    \item \textbf{Agent position and rotation}: the agent’s position and rotation in the simulation world; 
    \item \textbf{Messages}: the messages sent by all the agents.
\end{itemize}

\paragraph{Action Space} There are 7 types of actions for agents to interact with the environment or communicate with each other. Each action takes several frames and the detailed action space is listed here:
\begin{itemize}
\item \texttt{Move forward}: move forward 0.5m;
\item \texttt{Turn left}: turn left by 15 degrees;
\item \texttt{Turn right}: turn right by 15 degrees;
\item \texttt{Grasp}: grasp an object, only the agent is close to the object can he perform the action successfully. The object can be either a target or a container;
\item \texttt{Put in}: put the target into the container, only the agent is holding a target in one hand and a container in another hand can he perform the action.
\item \texttt{Drop}: drop the objects held in hand;
\item \texttt{Send message}: Send a message to other agents. In each frame, no more than 500 characters can be sent.
\end{itemize}

\subsubsection{C-WAH}

\paragraph{Tasks} C-WAH features five household task scenarios: Prepare afternoon tea, Wash dishes, Prepare a meal, Put groceries, and Set up a dinner table. Each task represents a different domestic activity and consists of 3-5 subgoals expressed as predicates in the format ``\texttt{ON/IN}(x, y)'' (meaning ``\texttt{Put} x \texttt{ON/IN} y\''). \cref{tab: tasks_CWAH} provides detailed descriptions of each task.
Agents must satisfy all specified subgoals within a 250-step time limit to successfully complete a task.

\begin{table}[ht]
\centering
\caption{Tasks in C-WAH.}

\label{tab:predicate_set}
\begin{threeparttable}
\label{tab: tasks_CWAH}
\begin{tabular}{p{4cm}p{8cm}}
\toprule
\textbf{Task Name} & \textbf{Predicate Set} \\ \midrule
Prepare afternoon tea & 
\texttt{ON}(cupcake, coffeetable), \texttt{ON}(pudding, coffeetable), 
\texttt{ON}(apple, coffeetable), \texttt{ON}(juice, coffeetable), 
\texttt{ON}(wine, coffeetable) \\ \midrule
Wash dishes & 
\texttt{IN}(plate, dishwasher), \texttt{IN}(fork, dishwasher) \\ \midrule
Prepare a meal & 
\texttt{ON}(coffeepot, dinnertable), \texttt{ON}(cupcake, dinnertable), 
\texttt{ON}(pancake, dinnertable), \texttt{ON}(poundcake, dinnertable), 
\texttt{ON}(pudding, dinnertable), \texttt{ON}(apple, dinnertable), 
\texttt{ON}(juice, dinnertable), \texttt{ON}(wine, dinnertable) \\ \midrule
Put groceries & 
\texttt{IN}(cupcake, fridge), \texttt{IN}(pancake, fridge), 
\texttt{IN}(poundcake, fridge), \texttt{IN}(pudding, fridge), 
\texttt{IN}(apple, fridge), \texttt{IN}(juice, fridge), 
\texttt{IN}(wine, fridge) \\ \midrule
Set up a dinner table & 
\texttt{ON}(plate, dinnertable), \texttt{ON}(fork, dinnertable) \\ \bottomrule
\end{tabular}
\end{threeparttable}

\end{table}

\paragraph{Observation Space} Agents in C-WAH receive these observations:
\begin{itemize}
    \item \textbf{RGB image:} Forward-facing egocentric view with $256 \times 512$ resolution and $60^{\circ}$ field of view;
    \item \textbf{Depth image}: Depth information with identical camera parameters as the RGB image;
    \item \textbf{Oracle Perception} (optional): Color-coded object segmentation map where each object ID corresponds to a unique color, using the same camera configuration;
    \item \textbf{Agent position}: Coordinates indicating the agent's location in the simulation environment;
    \item \textbf{Messages}: Communication messages received from all agents in the environment.
\end{itemize}

\paragraph{Action Space}  There are 8 types of actions for agents to interact with the environment or communicate with each other. Each action takes several steps and the detailed action space is listed here:
\begin{itemize}
\item \texttt{Walk towards}: Navigate to an object within the agent's current room or move to another room;
\item \texttt{Turn left}: Rotate counterclockwise by $30^{\circ}$;
\item \texttt{Turn right}: Rotate clockwise by $30^{\circ}$;
\item \texttt{Grasp}: Pick up an object, executable only when the agent is nearby;
\item \texttt{Open}: Open a closed container, executable only when the agent is nearby;
\item \texttt{Close}: Close an open container, executable only when the agent is nearby;
\item \texttt{Put}: Place held objects into an open container or onto a surface, executable only when the agent is close to the target location;
\item \texttt{Send message}: Transmit a communication (maximum 500 characters) to other agents.
\end{itemize}

\subsection{Experimental Infrastructure}  \label{supp: infrast}

The experiments were conducted using NVIDIA A100 GPUs. For LLaMA 3.1-70B APIs, we used the service provided by the TogetherAI platform. For ChatGPT-4o APIs, we used the service provided by OpenAI. Each experimental run required less than 2 days to complete.

\section{More Experimental Results} \label{supp:appendix_more_results}

\begin{figure}[h]
\begin{center}
\centerline{\includegraphics[scale=0.65]{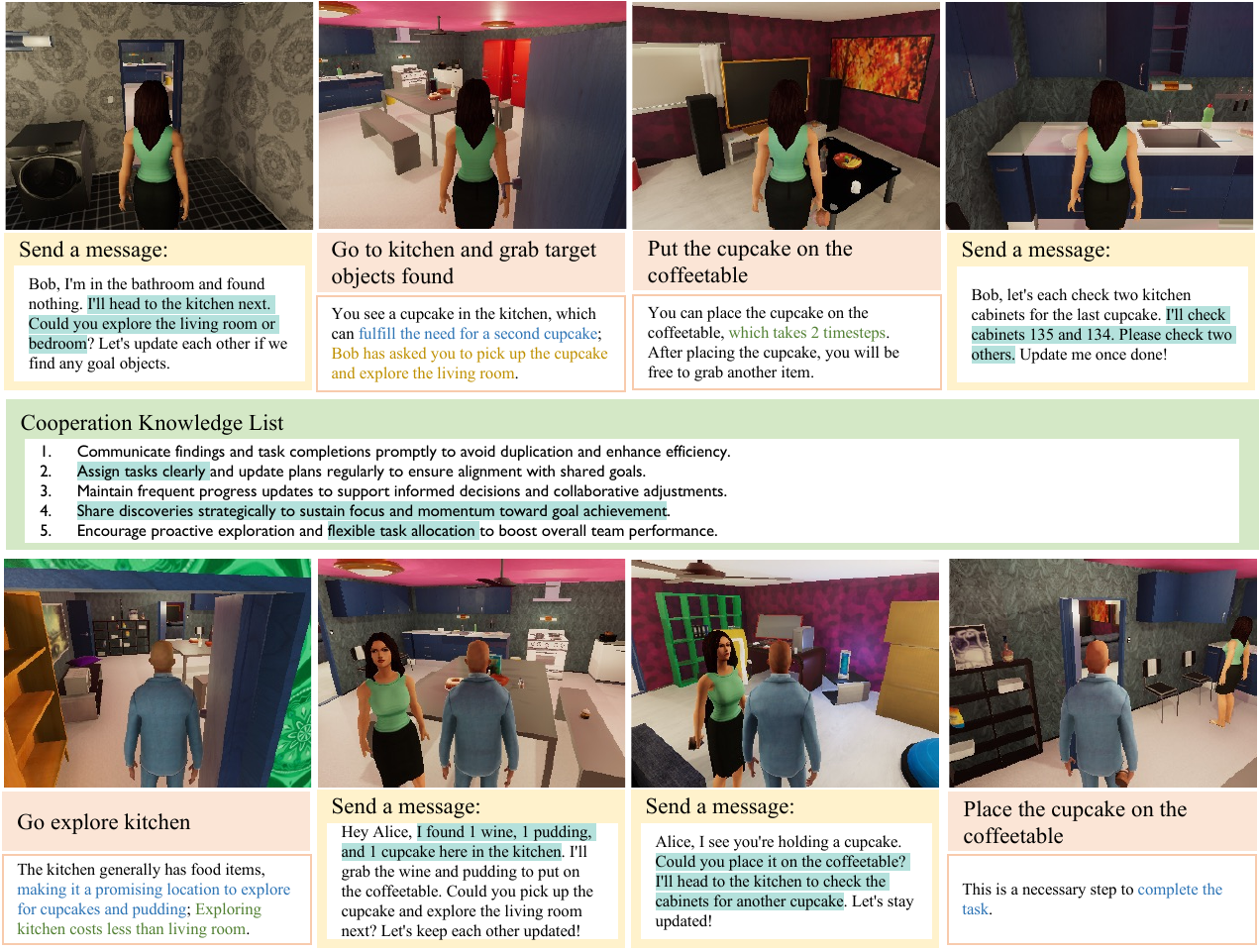}}
\caption{Snapshots of an example episode from the C-WAH benchmark with the task: ``Find and put 1 wine, 2 cupcakes, 1 pudding onto the <coffeetable>''. Each agent’s visualization includes annotations of their plans and thoughts. The current cooperation knowledge list is also displayed. The message
in the yellow dialogue box is directly generated by the LIET agent, while the thoughts are compact excerpts from the agent’s Chain of Thought (CoT) process.}
\label{fig:vis_comm}
\end{center}
\end{figure}

In this section, we provide additional visualization results of the proposed LIET policies in \cref{fig:vis_comm}. We observe that the decision process in this visualized episode comprehensively integrates \textcolor[rgb]{0.2588,0.4510,0.6941}{task progress}, \textcolor[rgb]{0.3686,0.5059,0.2471}{efficiency}, and \textcolor[rgb]{0.7216,0.5725,0.1882}{dialogue history}, further validating the conclusions presented in \cref{sec:exp_results}. In this particular episode, since both agents are concentrated in the kitchen and living room (where most target objects are located) during the early stages, we observe more intense discussions regarding labor division and progress updates. Notably, the knowledge list hints that agents should ``assign tasks clearly'' and ``share discoveries strategically to sustain focus and momentum toward goal achievement''. This guidance leads to explicit task division proposals during communication. At the beginning of the episode, Alice suggests division at the room level, but as the episode progresses, the proposed division becomes increasingly specific—ultimately reaching the object level by the end, in alignment with the knowledge list's recommendation for ``flexible task allocation''. This cooperative strategy exemplifies LIET's ability to foster effective collaboration through adaptive communication patterns.

\end{document}